\documentclass[10pt]{article}

\usepackage{graphics}
\usepackage{graphicx}

\usepackage[a4paper, left=35mm,right=35mm,top=34mm,bottom=34mm]{geometry}
\usepackage[utf8]{inputenc}
\usepackage[T1]{fontenc}
\usepackage[english]{babel}

\usepackage{enumerate}
\usepackage{graphicx}
\usepackage{hyperref}
\hypersetup{
    colorlinks=true,
    linkcolor=blue,
    filecolor=magenta,      
    urlcolor=cyan,
}

\usepackage{mathtools,amsthm,amssymb,amsfonts}
\usepackage{algorithm}
\usepackage{algorithmic}
\makeatother
\theoremstyle{plain}

\theoremstyle{definition}

\theoremstyle{remark}

\usepackage{caption} 
\usepackage{listings}
\usepackage{color}
\definecolor{dkgreen}{rgb}{0,0.6,0}
\definecolor{gray}{rgb}{0.5,0.5,0.5}
\definecolor{mauve}{rgb}{0.58,0,0.82}
\usepackage{fancyhdr}

\author{
  {\normalsize Fr\'ed\'eric Magoul\`es}\thanks{CentraleSup\'elec, Universit\'e Paris-Saclay, France.
    (correspondence, frederic.magoules@hotmail.com).}
  \and
  {\normalsize Qinmeng Zou}\thanks{CentraleSup\'elec, Universit\'e Paris-Saclay, France.}
}
\title{GPU Accelerated Contactless Human Machine Interface for Driving Car}
\date{}

\begin{document}
\maketitle
\thispagestyle{fancy}

\begin{abstract}
\noindent In this paper we present an original contactless human machine interface for driving car.
The proposed framework is based on the image sent by a simple camera device, which is then processed by various computer vision algorithms.
These algorithms allow the isolation of the user's hand on the camera frame and translate its movements into orders sent to the computer in a real time process.
The optimization of the implemented algorithms on graphics processing unit leads to real time interaction between the user, the computer and the machine.
The user can easily modify or create the interfaces displayed by the proposed framework to fit his personnel needs.
A contactless driving car interface is here produced to illustrate the principle of our framework.
\end{abstract}


\section{Introduction}

Nowadays, there is not a domain where people don't try to remove wires from computer devices.
So, controlling machines at distance, without any physical contact between the human and any object in communication with the machine, is more than what we can hope.
Reviews on human machine interaction based on computer vision can be found in~\cite{cipolla1998cvh,jr99survey,moeslund06}.
Within these systems, contactless human machine have been developed such as hand gloves~\cite{sturman1994sgb} or other tracking devices for human motion capture.
Next to wireless mouses and keyboards, technologies such as the virtual keyboard or the WII appeared.
Commercialized at the end of the year 2006, the Nintendo' console belongs to the new generation of human-machine interface.
The gamepad is the real innovation of this console.
Called Wiimote by Nintendo, the gamepad is indeed like a remote control, connected to the console by the Bluetooth technology.
But the Wiimote is more than a simple gamepad.
Thanks to an accelerometer, it is able to capture the player's hand's movements: left-right, up-down, forward-backward, rotation, etc.
It also incorporates a pointer device, like a computer mouse.
And, of course, it can also be used as a normal gamepad when it is horizontally handed.
The console offers other high-tech functionalities, while being compatible with former consoles (such as NES, SuperNES, Nintendo64, etc.).
It gives, between other things, the possibility of playing directly online.
This console is mainly destined for entertainment, and behind the gaming aspect, it brings the possibility of making sport without going out one's room, with the WIISports (Box, Bowling, Golf, Tennis, BaseBall, etc).

With one of our former paper, we have presented previously a framework where the number and the cost of equipment are significantly reduced: only a basic camera is needed, although the same functionalities are available, while opening new perspectives.
This framework is an open source framework that enables contactless human machine interaction using computer vision techniques.
This framework enables to implement any algorithm requiring a human-machine interaction, without any physical contact.
It is developed in C++ and is written in an object oriented approach.
Open-MP library and MPI library are used for parallel computations on CPU, and CUDA and OpenCL languages are used for parallel computing on GPU.
OpenCV is used for the image processing kernel.

In this paper we present how we have extended this framework for contactless human machine interface for driving car with graphics processing unit accelerated computations.
The plan of the paper is the following.
Section~\ref{sec:2} presents the architecture of the framework.
Section~\ref{sec:3} discusses some implementations issues of the algorithms.
Section~\ref{sec:4} illustrate an experiment for the driving.
Finally, Section~\ref{sec:5} concludes this paper.

\section{Architecture}
\label{sec:2}

Our framework works in two steps.
The first one consists in the initialization.
An XML file is read, containing the configuration to use.
Then, the FIZI module learns the background without the user, by using some advanced machine learning techniques.
Finally, the interface XML file is loaded.
The second step consists in the running one.
The camera gets an image, that is treated to deal with the hand's movement.
A first treatment is applied to the image so that it can be treated by the program.
The brightness is also corrected.
Then, the FIZI module can isolate the skin color zones, and more particularly, the hand's one.
This hand zone is converted into a pointer by the Mouse module.
The Interface module can use this pointer to get its movement, its state (click or not).
After this step, it applies the actions associated via the Engine module.
These four modules are shortly described in the following paragraphs: FIZI module, Mouse module, Interface module and Engine module.

\paragraph{FIZI module}
Its aim is to isolate the skin color zones.
Therefore, it implements the settings of the luminance and the contour detection parameters, which are easily adjustable to the functioning conditions.
After learning the background at the starting of the program, it can isolate a hand.

\paragraph{Mouse module}
Getting back the binary image from the FIZI module (skin color zones or not), this module tracks the hand corresponding zone.

\paragraph{Interface Module}
On the two previously mentioned modules, can be added the Interface module of which the role is to load a XML configuration file.
This file implements a graphic interface in accordance to the wanted application, linking zones to events.
With the Mouse module, it determines the interface zone activated by the pointer.
The interface is based on the OpenCV library which makes it independent from the Operating System on which it runs.

\paragraph{Engine module}
That last module is able to interpret the actions launched by the Interface module, in order to launch system's actions.
Opposite to the Interface module, the module is OS dependent.

\begin{figure}
  \centering
  \includegraphics[angle=0,trim={60 0 0 0},clip=true,width=0.8\textwidth]{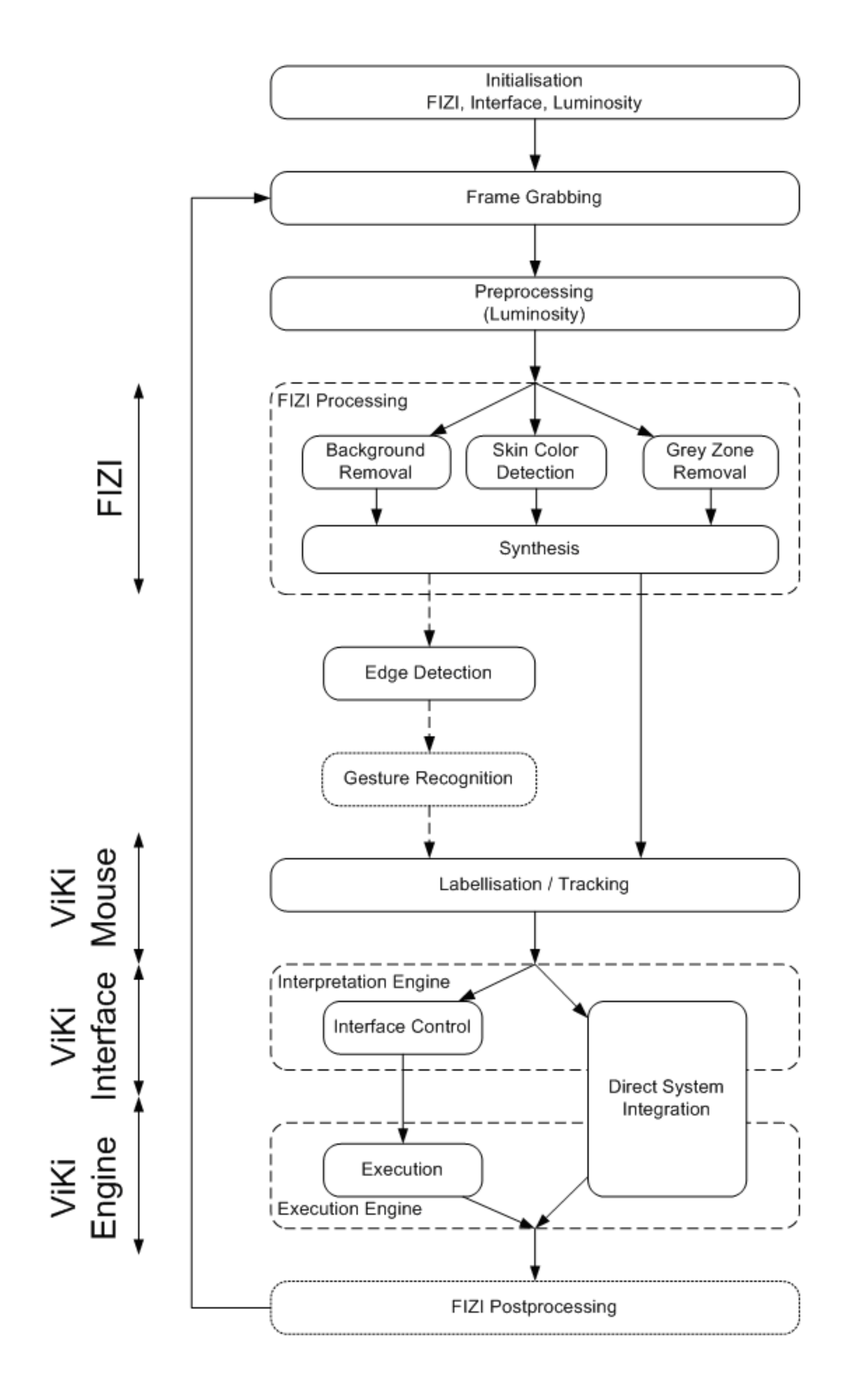}	
  \caption{Architecture of the framework.} \label{fig:framework}
\end{figure}

\paragraph{Workflow}
A complete description of the workflow is illustrated in Figure~\ref{fig:framework}.
Our framework can be adapted for any particular application.
The simplest way is to modify only the XML file.
But this enables only to adapt the graphical interface and to launch only some basic actions on some events defined with the interface.
The other way enables to implement more complex algorithm while modifying directly the main function.

\section{Implementation}
\label{sec:3}

\subsection{Building the Mask of the Hand}

\begin{figure}
  \centering
  \includegraphics[scale=0.4]{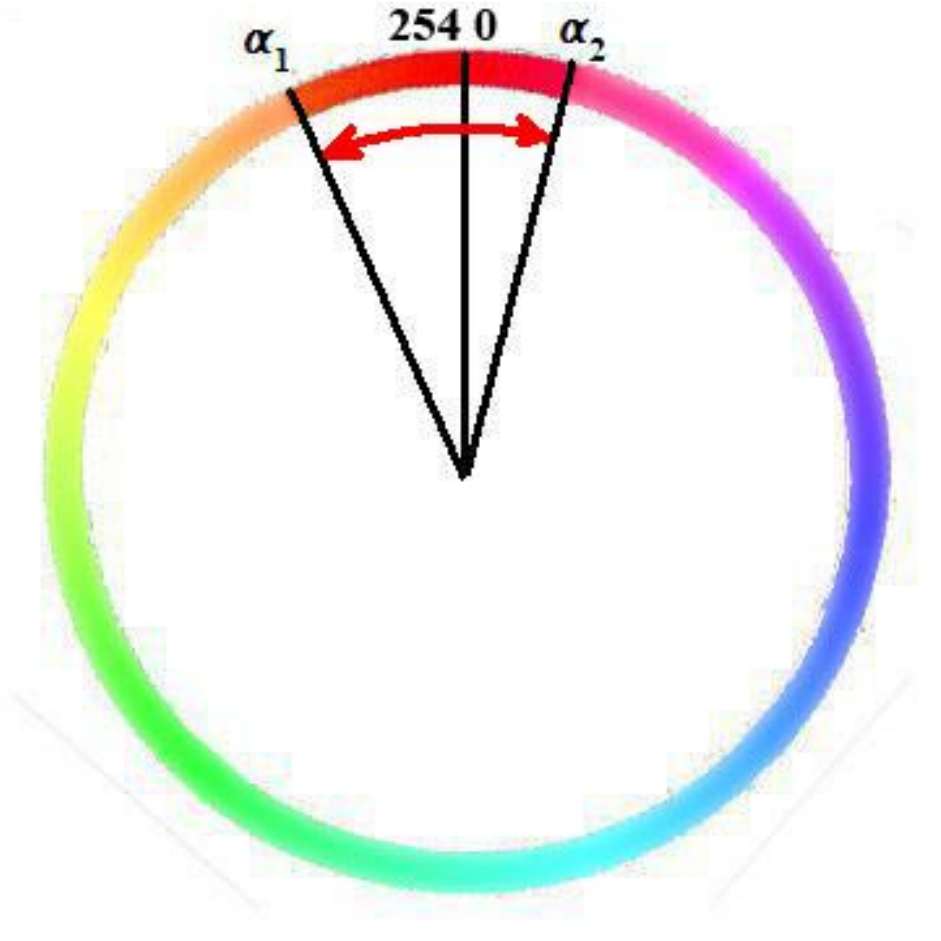}
  \caption{Threshold values on the Hue space.}
  \label{fig:hueseuil}
\end{figure}

To build the mask efficiently with respect to the need of low computational time, FIZI runs three different images in parallel.
These three images are the same image captured by the camera but in different color spaces:
(i) An RGB image ($I_1$) is used to remove the background using the learned parameters.
(ii) Another RBG image ($I_2$) is used to remove the gray zones.
(iii) An HSV image($I_3$) is used to detect and to segment the skin zones in the image.
For each of the three images, the processing corresponds to a basic threshold processing, as follow.
\\
(i) In image $I_1$, pixels which are between the minimum and the maximum values learned in the initialization step are removed and all the others are kept.
The resulting image $I_{R_1}$ is a binary image where white pixels represent the foreground and black pixels the background.
This corresponds to the values, $I_{R_1}(x,y)=1$ if $\min(I_{{bg}_{\min}}(x,y)) < I_1(x,y) < \max(I_{{bg}_{\max}}(x,y))$, $I_{R_1}(x,y)=0$ otherwise.
\\
(ii) In image $I_2$, an interactive threshold operator is applied.
For each pixel $(x,y)$ of the image $I_2$, if the difference between the minimum $\min_{RGB}$ and the maximum $\max_{RGB}$ values of the RGB components are lower than a tolerance $S$ defined by the user, the pixel is considered as a gray color pixel and thus is eliminated from the mask.
Indeed with high or low luminosity such pixels are often seen as undefined since their colors are arbitrarily set by OpenCV to zero value, i.e., red color, which unfortunately is included in the range of the skin's colors of European people.
This corresponds to the values, $I_{R_2}(x,y)=1$ if $\mid(\min_{RGB}(I_3(x,y)) - \max_{RGB}(I_3(x,y))\mid$, 
$I_{R_2}(x,y)=0$ otherwise.
\\
(iii) In image $I_3$, by using the color value in the HSL color space, the zones with color close to the hand's one can be isolated.
These pixels represent a mask of skin zones.
For the skin segmentation, two thresholds $\alpha 1$, $\alpha 2$ have to be defined on the Hue space as illustrated in Figure~\ref{fig:hueseuil}.
Defaults values have been learn from experiments but the end user can interactively modified them for his convenience and for the customization of the system.
These two thresholds are then used to extract the skin zones in images in the following way.
For each pixel $(x,y)$ of $I_3$, this corresponds to the values: $I_{R_3}(x,y)=1$ if $Hue(I_3(x,y)) \in [ \alpha_1, \alpha_2 ] \% 2 \pi$, $I_{R_3}(x,y)=0$ otherwise.
\\
At last, a merging step is processed to build the output meaningful result image.
The resulting mask is obtained by applying a logical AND operator between the three images.
Some morphological operations, combining erosion and dilatation operators are applied to remove the
small noisy objects and to connect to neighborhood zones.
As the output of FIZI, an image $I_{FIZI}$ represents the set of the hand zones of the end user.

\subsection{Managing Luminosity}

\begin{figure}
  \centering
  \begin{tabular}{cc}
  \includegraphics[width=0.3\textwidth]{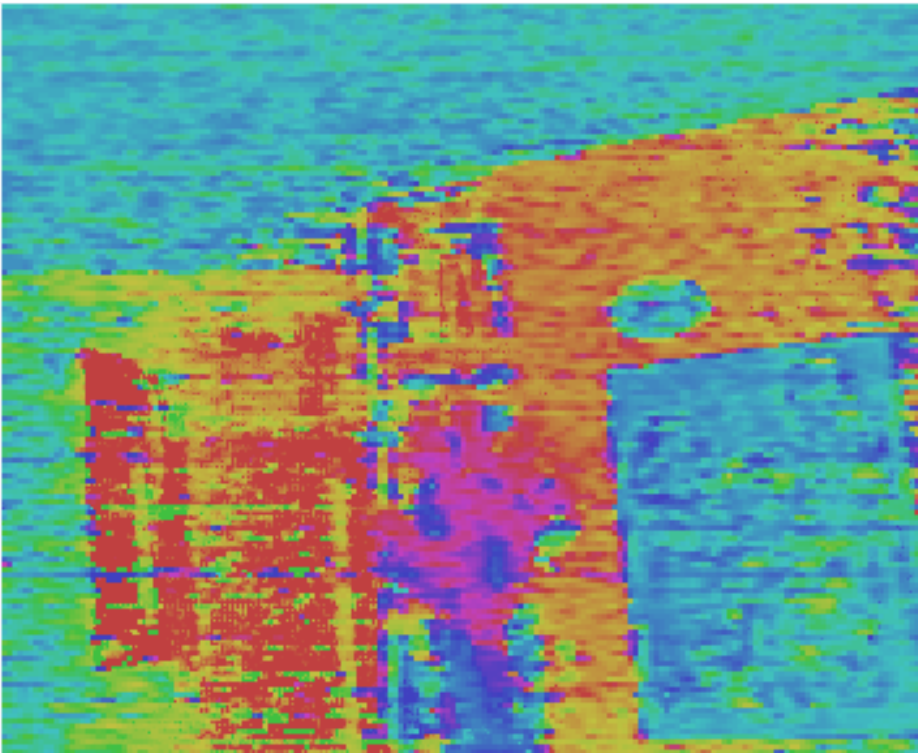} &
  \includegraphics[width=0.3\textwidth]{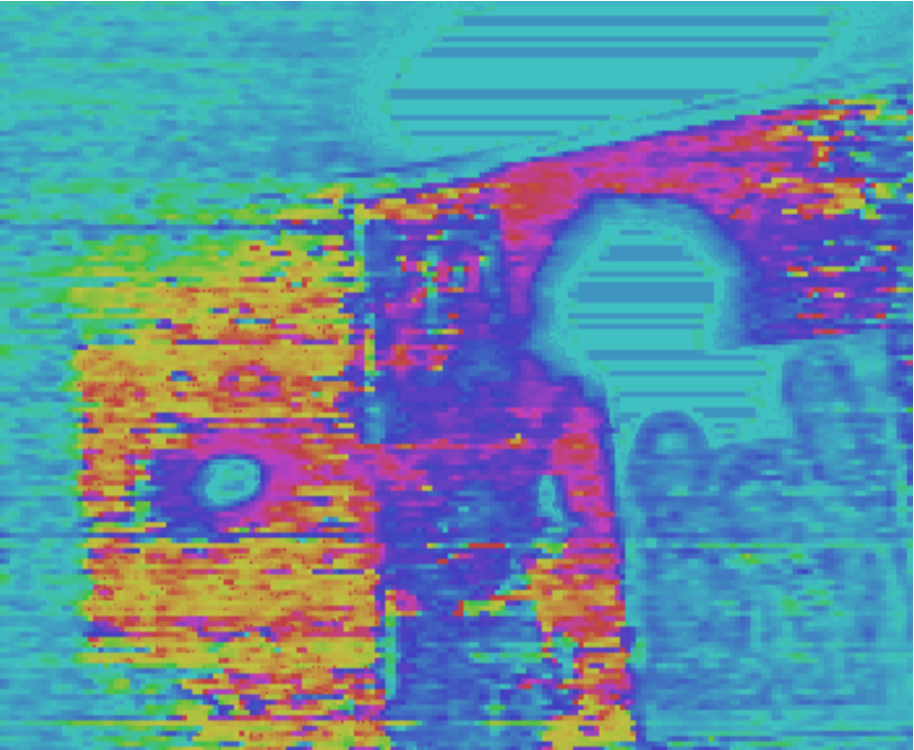}
  \end{tabular}
  \caption{Hand's color upon the luminosity.
  On the left, the hand looks violet and on the right two minutes after, the same hand looks blue.}
  \label{fig:hand-color}
\end{figure}

It is the biggest issue of the framework, since thresholds depends on the luminosity of the image.
While the application is running, the external conditions are subject to change (dawn or simply clouds modify the quantity of light received by the camera).
Since most camera adjust automatically, this leads to brutal changes within exposure time, in luminosity and colors.
To correct this we can apply an adaptive threshold or modify the dependent values consequently, as illustrated in Figure~\ref{fig:hand-color}.

\begin{figure}
  \centering
  \includegraphics[width=0.3\textwidth]{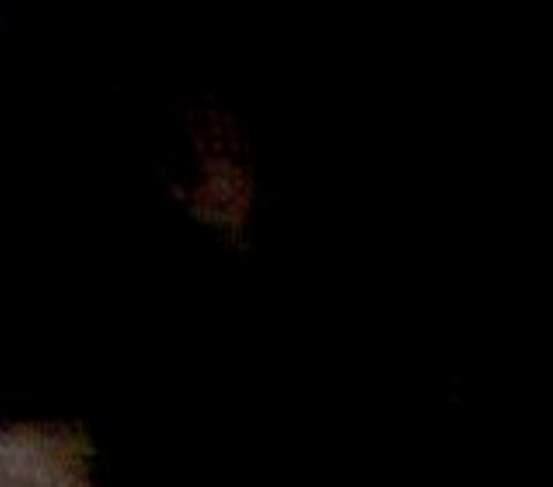}
  \includegraphics[width=0.3\textwidth]{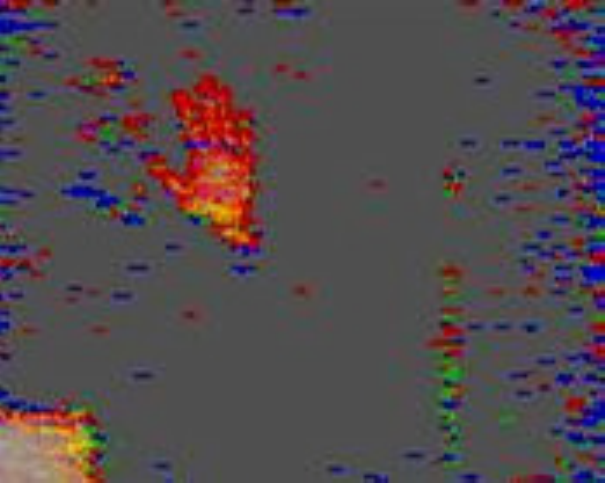}
  \caption{Luminosity treatment.} \label{fig:luminosity}
\end{figure}

Anyway, the automatic customization can trigger discontinuities in the luminosity leading to mistakes in the hand detection.
Luminosity raises issues in the process of the hand isolation by shifting the range of values in the different steps.
For instance, in a over-exposed environment, the hand is seen by the camera as a white zone but gray zones are removed during the process and thus the hand is not found.
A filter has been introduced to modify the Luminosity component in the most obvious cases, as shown in Figure \ref{fig:luminosity} (on the left, the image grabbed from the camera, on the right the image which can then be used as an input for the process).

\subsection{Acceleration with Graphics Processing Unit}

In addition of the existing modules, mathematical operations and linear systems solvers are implemented by Alinea (An Advanced Linear Algebra) library~\cite{magoules:journal-auth:58}.
The Alinea library is implemented in C++, MPI, CUDA and OpenCL.
For both central processing unit and graphic processing unit devices, there are different matrix storage formats~\cite{magoules:proceedings-auth:44,magoules:proceedings-auth:45}, and real and complex arithmetics in single- and double-precision.
It includes several linear algebra operations and numerous algorithms such as iterative methods~\cite{magoules:journal-auth:66,magoules:journal-auth:65}, domain decomposition methods in space~\cite{magoules:journal-auth:56}, together with some energy consumption optimization~\cite{magoules:journal-auth:61}.

It is well know that, dynamically reconfigurable vision-based user interfaces~\cite{icvs03} is a key issue to ensure efficient use of any framework.
In our framework, dealing with the evolution of the luminosity implies sometimes to re-initiate partly the machine learning techniques which improve the quality of the background removal~\cite{lee2005egm,Piccardi2004}.
Autotuning of the GPU memory is thus implemented~\cite{magoules:journal-auth:53,magoules:proceedings-auth:57,magoules:proceedings-auth:65,magoules:proceedings-auth:66} to improve the memory size and computation speed dynamically.

\section{Application for Driving Car}
\label{sec:4}

\begin{figure}
  \centering
  \includegraphics[angle=-90,trim={0 0 200 0},clip=true,scale=0.4]{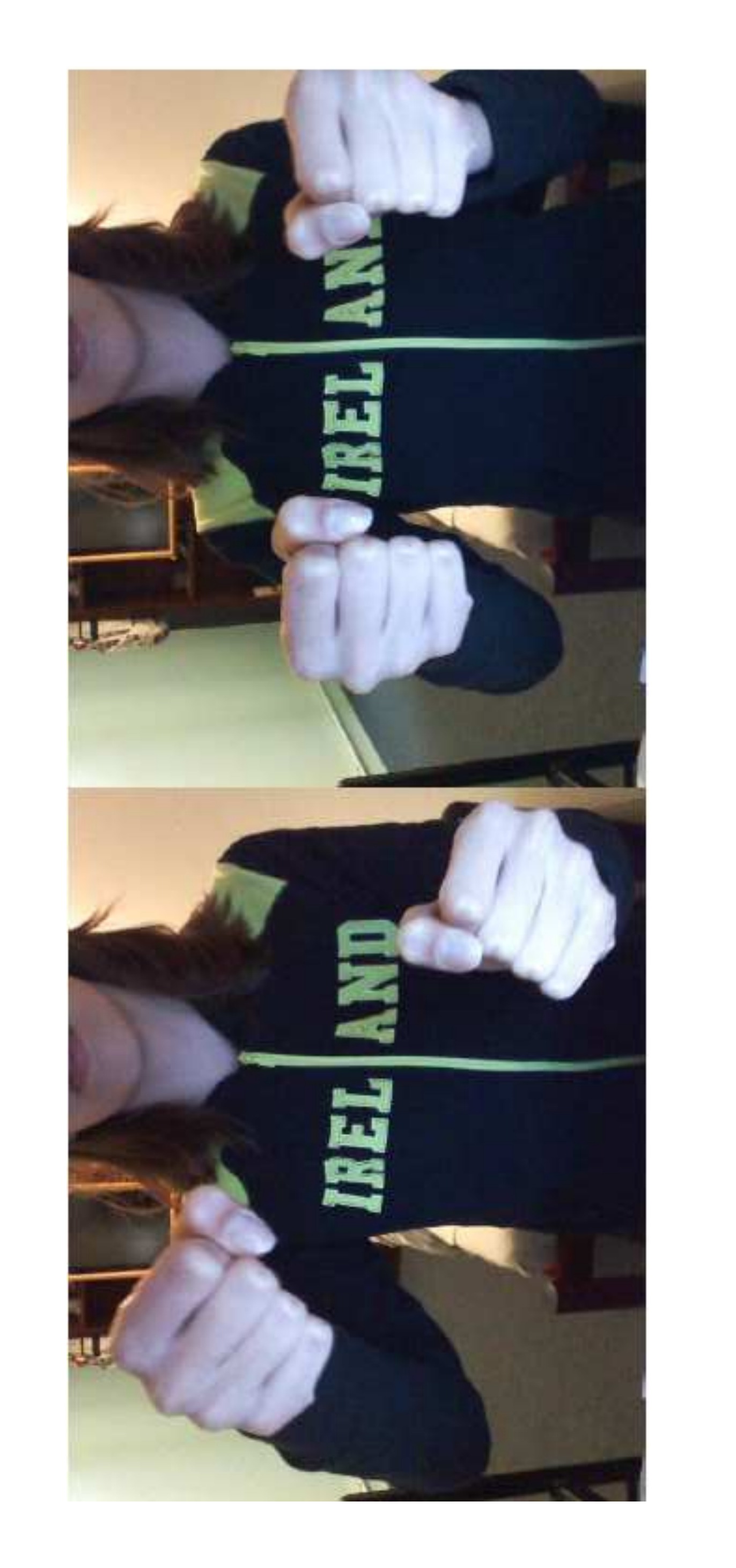}
  \includegraphics[angle=-90,trim={0 0 200 0},clip=true,scale=0.4]{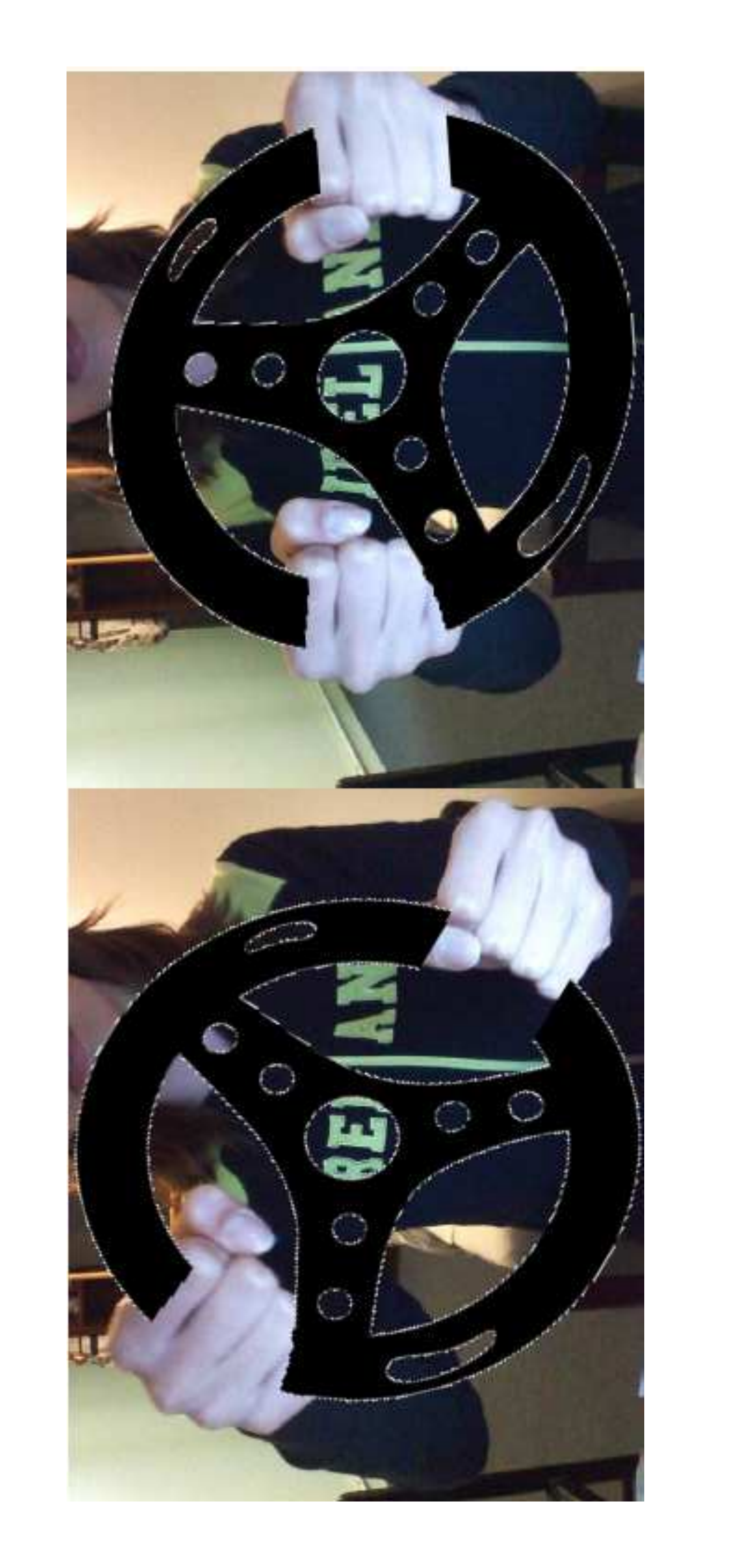}
  \caption{Contacless interface for driving car.} \label{fig:driving}
\end{figure}

Everybody has at least once played with a car driving game on the computer, probably with a console.
The algorithm behind mainly consists of controlling three objects: the left direction, the right direction and the speed of the car.
Within our framework, the movements of the hand consists on a rotation around an imaginary wheel.
To ease this rotation of the hand, an image of this wheel is added on the captured image by the camera to the screen, so the users knows where are his hand on the virtual wheel.
Figure~\ref{fig:driving} illustrates this interface.

\section{Conclusions}
\label{sec:5}

In this paper an original approach for driving car has been presented.
The proposed approach achieves a virtual wheel using an image acquisition device, for instance a camera.
Interaction of the user with the computer without physical contact by the only mean of a camera has been proposed.
Accurate luminosity detection and acceleration of the computations with graphics processing unit, make the proposed framework, a robust, accurate and real time human machine interface.
Application for driving car has been illustrated.

\section*{Acknowledgements}

The authors acknowledge the numerous students from Ecole Sup\'erieure des Sciences et Technologies de l'Ing\'enieur de Nancy (France) and from Ecole Centrale Paris (France) who have contributed to this framework since 2003, and in particular N. Vienne, J. Tavernier, the main programmers of the image processing workflow, J. Petin, N. Lambolez, M. Hjalmars, S. Cagnon, C. Mombereau, J. Ott, G. Mathias, S. Massot, D. Miliche, W. Ken, P. R\'emi, B. Rochet, J. Holburn, G. Sauwala, A. Brito Alves da Silva, A. Ortega, A. Vinicius Gonzalves Cardoso, F. Mirieu, P. d'Herbemont, E. de Roux, the main programmers of several modules, H.-X. Zhao, Sidharth GS, the main programmer of the machine learning techniques, and A.-K. Cheik Ahamed, K. Guilloy, M. Wang, M. Lubin, T. Stremplewski, E. Balagna, I. Petrovic, N. Foresta, N. Fleury, A. Roullier, B. Allardet-Servent, A. Desmaison, F. Mayer, H. Ben Salem, J.-C., T. Saint-Paul, T. Zhu, A. Manai, L. Beaubois, T. Brendl\'e, O. Darwiche Domingues, O. Soualhine, S. Beux, K. Towa, S. Ehrhardt, S. Olivotto, A. Bodin, M. Lagattu, M. Botarro, E. de Roffignac, C. Haroune, G. Cochet, the main programmers of GPU acceleration techniques.
Since 2006, this framework has been named ViKi (Virtual Interactive Keyboard Interface).

\bibliography{ref}
\bibliographystyle{abbrv}

\end{document}